\documentclass{vgtc}                          

\ifpdf
  \pdfoutput=1\relax                   
  \usepackage{graphicx}                
  \DeclareGraphicsExtensions{.pdf,.png,.jpg,.jpeg} 
\else
  \usepackage{graphicx}                
  \DeclareGraphicsExtensions{.eps}     
\fi%

\graphicspath{{figures/}{pictures/}{images/}{./}} 

\usepackage{microtype}                 
\PassOptionsToPackage{warn}{textcomp}  
\usepackage{textcomp}                  
\usepackage{mathptmx}                  
\usepackage{times}                     
\usepackage{cite}                      
\usepackage{tabu}                      
\usepackage{booktabs}                  

\usepackage[table]{xcolor} 
\usepackage{tabularx} 
\usepackage[margin=1in]{geometry} 

\definecolor{rowgray}{gray}{0.925}

\onlineid{0}

\vgtccategory{Research}

\preprinttext{Submitted to IEEE VIS Posters, 2023.}

\title{
Towards Metrics for Evaluating Creativity in Visualisation Design}

\author{Aron E. Owen\thanks{e-mail: aron.e.owen@bangor.ac.uk}\\ %
        \scriptsize Bangor University %
\and Jonathan C. Roberts\thanks{e-mail: j.c.roberts@bangor.ac.uk}\\ %
     \scriptsize Bangor University %
}

\authorfooter{
	\item J.C. Roberts, A.E.Owen and P.D. Ritsos are with the School of Computer Science and Electronic Engineering, Bangor University, UK. 
	E-mail: {j.c.roberts; aron.e.owen;}@bangor.ac.uk.}
 
\abstract{
Creativity in visualisation design is essential for designers and data scientists who need to present data in innovative ways. It is often achieved through sketching or drafting low-fidelity prototypes. However, judging this innovation is often difficult.  A creative visualisation test would offer a structured approach to enhancing visual thinking and design skills, which are vital across many fields. Such a test can facilitate objective evaluation, skill identification, benchmarking, fostering innovation, and improving learning outcomes. In developing such a test, we propose focusing on four criteria: Quantity, Correctness, Novelty, and Feasibility. These criteria integrate into a test that is easy to administer. We name it the Rowen Test of Creativity in Visualisation Design; We introduce the test, scoring system and results from using eight visualisation experts. 

} 

\CCScatlist{
  \CCScatTwelve{Human-centered computing}{visualisation design}{}{};
}

\begin{document}
\firstsection{Introduction}

\maketitle

Sketching and low-fidelity design are foundational in visualisation. They allow designers to rapidly iterate and explore concepts without the constraints of full system development. Quick and efficient sketches facilitate early identification of poor design choices, proving essential for brainstorming and crafting innovative visual solutions.
Despite their undeniable importance, there is currently no structured method to evaluate the creativity and quality of these sketches in the context of visualisation design. Our research focuses on the design and creation of novel visualisation ideas. When working on the VisDice innovation project~\cite{owen2023inspire} we needed a strategy to evaluate design sketches. Inspired by the Torrence test of creative thinking~\cite{torrance1966torrance} and from Guildford's diverse uses for a brick~\cite{guilford1956structure}. This paper introduces our test, discusses some related work, and provides an outline evaluation. We aim to expand and integrate the test into a broader creative visualisation design process.

\section{Related Work}
Guilford's~\cite{guilford1956structure} 1950's work revolutionised creativity assessment by focusing on divergent thinking. He measured fluency (quantity of ideas), flexibility (variety of categories explored), originality (uniqueness of ideas), and elaboration (depth of detail). His tasks included generating wild uses for everyday objects or imagining consequences of unlikely scenarios. Torrance~\cite{torrance1966torrance} built on this work with 
his Torrance Tests of Creative Thinking (TTCT). In visualisation, ideas of quantity are included as a measure of grading `good practice' in the first sheet of the Five Design-Sheet method~\cite{RobertsHeadleandRitsos16}. Where 10-15 ideas are a good indicator of creativity, while the visual literacy test (VLAT) provides a literacy score~\cite{LeeETALvlat2017}.
Unfortunately, these methods do not address the specific needs of visualisation design, which focuses on effectively representing data through innovative visual means. 
Other approaches, such as the ADDIE Model~\cite{Branch2009} and SCAMPER~\cite{Ozyaprak2015SCAMPER}, Tripartite Model~\cite{thrash2003inspiration}, Schneiderman Genex~\cite{ShneidermanCreatingCreativity2000}, and Wallas's work~\cite{Wallas1926art} offer structured thinking in creativity, but no successful metrics. While deBono~\cite{deBono2009six}, Meyer et al~\cite{Meyer} and Johnson~\cite{johnson2011good} offer ideation strategies but no evaluation method.  What is missing are strategies specific for measuring low-fidelity sketches and ideation in visualisation.

\section{Building the metrics and evaluation}
The absence of a dedicated evaluation framework for low-fidelity sketches in visualisation design leaves a significant gap in our ability to measure and understand creativity in this field. This gap hinders developing and refining effective visualisation solutions, as designers need more feedback to improve their ideas and techniques. Guildford's work highly influenced our ideas. His framework enhances our understanding of creativity and guides educational strategies and professional assessments, fostering the development of creative potential across diverse fields and demographics.
Our test (the Rowen test) aims to provide a similar comprehensive framework for visualisation, focusing on low-fidelity designs. 
We decided on four dimensions because we wanted to use only a few dimensions yet focus on relevance for visualisation. 
\textbf{Quantity} assesses the number of iterations or variations in the sketch, reflecting the breadth of ideation during the creative process.
\textbf{Correctness} evaluates whether the sketch clearly depicts the intended chart type and accurately represents the underlying data.
\textbf{Novelty} assesses the originality of the sketch by determining whether it introduces new and unique ideas or merely replicates standard visualisations. Furthermore, this dimension focuses on the creativity and innovation displayed in the visualisation design.
\textbf{Feasibility} evaluates whether the sketch is practical and can be realistically implemented using existing tools and techniques. This dimension considers the practicality and achievability of the proposed visualisation design within current technological and resource constraints.
We place the dimensions in a Likert (0..25), allowing a maximum overall value of 100. Moreover, this offers a structured approach to measuring creativity and quality in preliminary visualisation designs. We add a box and reflection to make a creative metric sheet. 

\begin{figure}[ht]
  \centering
  \includegraphics[width=\linewidth]{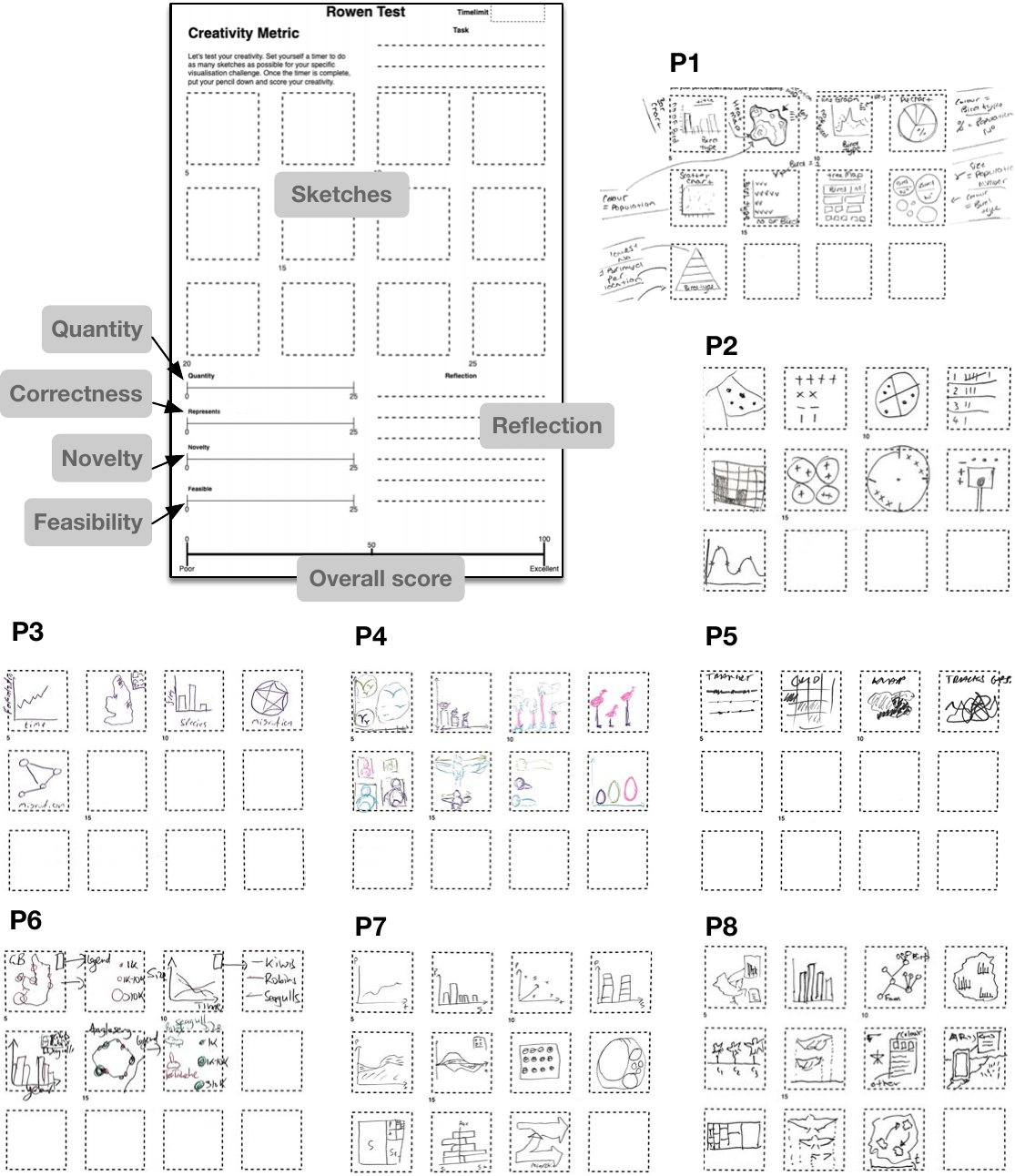}
  \caption{\label{fig:ddf} The Rowen test (top left) shows the eight participants. They self-assessed their work. Quantity is easy to judge, with P8 being the most and P5 being the least. Creativity is highest in P1, with P7 focusing on typical charts (including line graphs, bar charts, stacked bars and so forth).\vspace{-5mm} }
\end{figure}

After running a pilot study, we ran an evaluation (with ethics approval CSEE-R-2024-CG-003) with 8 experts in visualisation, who answered our call for volunteers. We asked them to imagine planning to visualise the bird migration data~\cite{Tobias2022}, sketch visualisation ideas, self-evaluate the four metrics, and answer the System Usability Scale (SUS) questions~\cite{bangor2008}, and express positive and negative comments.
Participants provided a range of feedback. They expressed positive aspects, such as the test being easy to use and quick to execute, which encouraged creativity (P2, P3). The simplicity and lack of cognitive complexity let participants quickly move to new ideas and iterate more interesting designs (P4). The hands-on nature of sketching was described as fun, and the post-scoring aspect prompted further reflection (P5, P6). Additionally, the test challenged participants to develop diverse sketches, pushing them to think outside the box (P8). Across the eight experts, a total of 64 images were created.
Some participants who made critical remarks felt they required additional context to assess creativity (P6) accurately. We appreciate their feedback and will consider these points for future iterations of the Rowen Test. 


\section{Discussion and conclusion}
A test can offer a holistic view of an individual's creative potential by assessing multiple dimensions. Furthermore, this approach can inform best practices in visualisation design, ensuring outputs are aesthetically appealing and practically insightful, thereby advancing theoretical knowledge and practical applications in visual communication and data visualisation.
The positive feedback underscores the test's effectiveness in fostering creativity through its simplicity and hands-on approach. Participants appreciated the test's quick and easy-to-use nature, which allowed them to iterate rapidly and generate diverse ideas. Moreover, this aligns with the fundamental goal of low-fidelity sketching: to quickly and roughly explore a wide range of concepts without being constrained by high-fidelity tools, a goal that the Robwen Test supports. 
However, the evaluation also revealed areas for improvement. Specifically, participants expressed a need for clarification on the scoring system, which suggests that this aspect of the test needs refinement. This emphasis on the potential for improvement should make the audience feel hopeful about the test's future effectiveness. Additionally, the challenge of sketching without a comprehensive understanding of the data highlights the need to provide more context to participants. Addressing these issues could enhance the Rowen Test's ability to evaluate creativity more effectively.
Inspiration and creativity are complex and multifaceted subjects that extend beyond the scope of this study. The role of inspiration in fostering creativity, the influence of external stimuli, and the psychological processes underpinning creative thought are all critical factors that warrant further investigation. A detailed exploration of these themes would provide a deeper understanding of cultivating and measuring creativity, particularly in visualisation design.
This discussion opens the conversation on the broader implications of evaluating creativity in design. While the Rowen Test offers a structured approach to assessing low-fidelity sketches, it is clear that any single test or metric can only partially capture creativity. Future research should explore the intricate relationship between inspiration, creativity, and practical design constraints. Such investigations could lead to developing more comprehensive frameworks and tools for fostering and assessing creativity in various fields.

In conclusion, the Rowen Test represents a significant step forward in evaluating low-fidelity sketches. It provides a valuable framework for measuring creativity and highlights the need for ongoing refinement and deeper exploration of the underlying factors that drive creative thought. Furthermore, this stresses the importance of continuing the conversation, engaging the audience and emphasising their integral role in advancing our understanding of creativity and improving design practices.

\bibliographystyle{abbrv-doi}

\bibliography{new}
\end{document}